\documentclass[%twocolumn,
aps,prd,  
               preprintnumbers,numbers,sort&compress,
               nofootinbib, showpacs, colorlinks,
               linkcolor=blue, citecolor=blue]{revtex4-2} % reprint,nofootinbib

\usepackage{graphicx}% Include figure files
\usepackage{dcolumn}% Align table columns on decimal point
\usepackage{bm}% bold math
\usepackage{hyperref}
\usepackage{xcolor}

\usepackage{amsmath}
\usepackage[caption=false]{subfig}
\usepackage{gensymb}% to type the \degree symbol
\usepackage{amsmath}
 \usepackage{bm}% bold math
\usepackage{graphicx}% Include figure files
\newcommand{\be}{\begin{eqnarray}}
\newcommand{\ee}{\end{eqnarray}}
\newcommand{\beq}{\begin{equation}}
\newcommand{\eeq}{\end{equation}}
\newcommand{\exclude}[1]{}

\def\ra{\rangle}
\def\la{\langle}

 \newcommand{\jcap}{JCAP}
\newcommand{\mnras}{Mon.\ Not.\ R.\ Astron.\ Soc.}
 \graphicspath{{./Figures/}} 

\begin{document}
    
       \title{Mysterious Transients in the Palomar Observatory Sky Survey (POSS-1)    as   profound manifestation of  the Dark Matter physics }
        \author{Ariel  Zhitnitsky  }
               \email{arz@phas.ubc.ca}
       \affiliation{  University of British Columbia,Vancouver,  Canada}

     \begin{abstract}
     {Transient star-like objects of unknown origin have been identified in the first Palomar Observatory
Sky Survey (POSS-1) as part of the Vanishing and
Appearing Sources during a Century of Observations (VASCO) project. The source of the transients recorded by POSS-1    remains unknown, which is the warrant  to coin the observed phenomena as Mysterious Transients (MT).   We advocate an idea that  the dark matter  (DM) in form of the     axion quark nuggets (AQN)    made of standard model  quarks (or antiquarks) and gluons, similar to the       old idea of the Witten's strangelets,  could 
{\it simultaneously} explain {\it all }  the observed MT signals (including very short time scale for flash itself, association with nuclear test timing, observed alignments of several MT events, correlation with UAP reports, etc)  collected or recorded for many years.
 Essentially we argue that the MT is a cousin of  Ball Lightning (BL) events, also observed for centuries, without commonly accepted physics explanation.      The AQN dark matter model was invented long ago without any intension  to apply the model to MT or to BL phenomena. Rather,   it was invented with  a single motivation to explain the observed  similarity     $\Omega_{\rm DM}\sim \Omega_{\rm visible}$ between  visible and DM components, which  represents a very  generic feature of this framework,  not sensitive  to any parameters of the construction.  The basic parameters of this model (such as the typical baryon charge of the nuggets) 
 had been fixed long ago by explaining the observed excess of radiation at variety of scales: from galactic to the solar, to local Earth's environments.  In this work we use the same framework with the same set of parameters to study the observed MT phenomena.  We also suggest several tests  which substantiate or refute our proposal. We also present   some  suggestions on   type of instruments required to  study this specific (and well defined) type  of the   UAP events representing  the  cousins of BL and MT events in the AQN framework.}
 
           \end{abstract}

	% \begin{keywords}
%	 dark matter,    axion quark nugget,  mysterious transients, ball lighting, sky-quakes, UAP  
%  \end{keywords}

\maketitle

\section{Introduction}\label{sec:introduction}

The title of this work seemingly includes two contradicting terms: the first one is  ``Mysterious Transients (MT)", 
which are very bright luminous  phenomena, though of  unknown nature. The second term of the title is  ``dark matter" (DM)  which is, by definition, must decouple  from the  radiation, as it cannot emit light. 
Another comment on the title: we use words ``profound manifestations of the Dark Matter physics"  in the present work devoted to MT events which is precisely the wording we used in our paper \cite{Zhitnitsky:2025bvy}
 devoted to Ball Lightning (BL) phenomenon titled ``Ball Lightning as a Profound Manifestation of Dark Matter Physics". This is not an accidental  coincidence in similarity of the titles, of course. Rather,    it is purposely   done
 to emphasize that     both phenomena are in fact close cousins as both deeply rooted to the same axion quark nugget (AQN)  framework when  the  DM  is represented by composite macroscopically large AQN objects made of the  standard model  quarks (or antiquarks) and gluons, in contrast with conventional 40 years old DM paradigm  when   DM particles   are  represented by some fundamental new (unknown and not observed) fields  of BSM (Beyond Standard Model) physics. 
 
Therefore, it is important to make a comment from the start    that the AQN   behaves as a chameleon: it serves as a proper DM object in dilute environment in empty Universe, but becomes very strongly interacting (with surrounding material) object 
 when it hits the stars or planets. Therefore,  the  contradiction in the title is only apparent as the DM in form of the AQNs   become strongly interacting objects in dense environment and can indeed produce  profound and very powerful events  such as MT,
 which is the topic of the present work.  

 \subsection{Brief overview of Mysterious Transient (MT) events}\label{items}
 The first   term  of the title is ``Mysterious Transient" events\footnote{We use term ``Mysterious Transient (MT)" in this work  to be consistent with  wording  being used in the original paper  \cite{Villarroel:2021ddh} ``... mystery of simultaneous transients is a detective story..."
.}.  Therefore, we start with brief overview of the basic features of the MT events as   identified by the  Palomar Observatory
Sky Survey (POSS-1) as part of the Vanishing and
Appearing Sources during a Century of Observations (VASCO) project \cite{2022MNRAS.515.1380S,Villarroel:2019bky,Villarroel:2020knw,Villarroel:2021ddh,2024MNRAS.527.6312S,Villarroel_2025}:

1. Mysterious Transient (MT) star-like objects are short lived (less than 30 min of exposure) objects. They cannot be related to  artificial
satellites as the  sky surveys had been conducted before 1957. Furthermore,  all known astrophysical explanations had been considered but  ruled out \cite{Villarroel:2021ddh,Villarroel_2025}. The recorded  magnitude in optical bands is  $M_{AB}\in (17-19)$ for MT events;

2. MT events  are sharper
and more circular than the images of stars. The MT events must be sub-second flashes of light \cite{Villarroel:2025xym,2026arXiv260320407B}.
Furthermore, slight asymmetries in the light profile are present in few cases, manifesting as mild elongation in shape \cite{Villarroel_2025}; 

3. MT events sometime  constitute several point like-transients  that are aligned along a line within a single exposure \cite{Villarroel:2021ddh,Villarroel_2025,Villarroel:2025xym}. What is the source for such very unlikely alignment? One should mention that the probability of two such transients appearing in the same $(10\rm ~ arcmin)^2$ box is $p\sim 10^{-6}$   \cite{Villarroel_2025}, and probability  for aligned events within a single exposure should be even much  smaller.

4. MT are strongly correlated with nuclear testing \cite{Bruehl:2025},  \cite{2026arXiv260400056D}. This is absolutely astonishing finding which is very hard to accommodate within conventional physics; 

5. MT are also strongly associated with Unidentified Anomalous Phenomena (UAP) reports  \cite{Bruehl:2025};

    \subsection{Brief overview of dark matter (DM) in the AQN framework}
   The second  term  of the title is ``dark matter". 
    Therefore, we have to briefly  explain  the term  ``dark matter".  From cosmological viewpoint there is a fundamental difference between dark matter
  and ordinary matter (aside from the trivial difference
 dark vs.  visible). Indeed, 
 DM played a crucial role in the formation of the present  structure in the universe.  Without dark matter, the universe would have remained too  uniform to form the galaxies.  
 Ordinary matter could not produce fluctuations to create any significant  structures   because it remains tightly coupled to radiation, preventing it from clustering, until  recent epochs.   
  The key parameter which enters all the cosmological observations is the corresponding cross section $\sigma$ 
  (describing coupling of DM with standard model particles) to mass $M_{\rm DM}$ ratio which must be sufficiently small
to play the role of the DM as briefly mentioned above,  see e.g. recent review  \cite{Tulin:2017ara}:
\be
\label{sigma/m}
\frac{\sigma}{M_{\rm DM}}\ll  1\frac{\rm cm^2}{\rm g}.
\ee
The  Weakly Interacting Massive Particles (WIMP) obviously satisfy to the criteria (\ref{sigma/m}) to serve as DM particles due to their very tiny  cross section $\sigma$ for  a typical mass $M_{\rm WIMP}\in( 10^2-10^3) ~\rm GeV$.
However, the WIMP miracle   
%which has been the dominant idea  for the last 40 years 
has  failed as dozen of dedicated instruments could not find any traces of WIMPs though the sensitivity of the instruments had dramatically improved by many orders of magnitude during the last decades. 

In the present work we consider a  fundamentally  different type of the DM which is in form of  
   macroscopically large composite objects of nuclear density, similar to the Witten's quark nuggets  \cite{Witten:1984rs,Farhi:1984qu,DeRujula:1984axn}.   The corresponding objects   are called the  AQNs and behave as    {\it chameleons}: they are (almost) 
    not interacting entities      in dilute environment, 
  such that the AQNs may serve as proper DM candidates as the corresponding condition (\ref{sigma/m}) is perfectly satisfied for the AQNs during  the structure formation when  the ratio $\sigma /M_{\rm AQN}\lesssim 10^{-10} {\rm cm^2}{\rm g^{-1}}$.  However, the same objects interact very strongly with material when they hit the Earth, or other planets and stars. The AQNs are absolutely stable objects on cosmological scales as the energy per baryon charge in AQNs is smaller than the energy per baryon charge for nucleons.  
  
The main distinct feature of the  AQN model (which plays absolutely crucial role for  the present work) 
in comparison with old Witten's construction 
 is that AQNs  can be made of {\it matter} as well as {\it antimatter} during the QCD transition as a result of the charge segregation   process, see original proposal \cite{Zhitnitsky:2002qa},  brief overview \cite{Zhitnitsky:2021iwg} and very recent lectures  \cite{VanWaerbeke:2026sxs} on the AQN framework with numerous applications at different scales, from   early Universe      to the galactic, solar  and Earth environments.

 The presence of the antimatter  nuggets in the system implies that there will be  annihilation events leading  to very profound  strong effects when antimatter AQNs hit the Earth's atmosphere.  
More specifically, 
 our claim here is that the  basic features of the MT events as listed by items 1.- 5. in Sect. \ref{items}
   may  naturally emerge  as a result of these  powerful and energetic   events   related to antimatter nature of the AQN objects.

     This model was invented long ago to resolve fundamental problems in cosmology, not related  to the topic of the present work, the MT events, see next section \ref{AQN} with details. Nevertheless, this AQN framework  may also shed a light on    the nature of the MT events as argued  in this work. We stress that all parameters for this model were
     fixed long ago  in  our previous applications  to  explain  a number of  mysterious and puzzling observations at the galactic, solar and Earth's scales.   We keep all these parameters of the AQN model identically the same to understand  the nature of the MT events.     These  parameters  had been extracted by  matching  the  computations in AQN model  with  observed   mysterious phenomena,   such  as  the puzzling  UV excess in our galaxy \cite{Sekatchev:2025ixu}.   
 
 The organization  of this work is   as follows. The two sections  \ref{AQN} and \ref{BL-features}   represent a brief overview of the AQN construction
 with emphasize on understanding of the BL phenomena which is the close cousin of the MT events in this framework. In our main Sect. \ref{sect:proposal}  we argue that the various of observations  as formulated  above  by items 1.- 5.  in Sect. \ref{items}  can be  naturally   explained within  AQN framework. In Sect.  \ref{sec:MT_event_rate} we estimate
 the MT event rate and find it is consistent with observations. In Sect. \ref{sect:cousins} we briefly mention on possible relation  of the  MT events with  their  other cousins (in addition to BL events mentioned earlier) such as  well recorded  pseudo-meteorites  events and sky-quakes, which can be classified as special type of the UAP events. In our concluding Sect. \ref{sect:tests} we suggest a number of specific tests which can substantiate  or refute our proposal on close relation between MT and dark matter physics within AQN framework. 
   
  \section{ The  AQN framework: the basics }\label{AQN}
  This section  serves as introductory 
  overview  of the fundamental  ideas of the AQN model.  
  %The  Sect.  \ref{BL-features}  serves as a brief overview of the original proposal \cite{Zhitnitsky:2025bvy}  when  all highly nontrivial features of the  BL events are explained within the   AQN  framework.  

% \subsection{The basics features of the AQN framework}\label{basics}
 The original motivation for the AQN model  can be explained in two lines as follows. 
It is commonly  assumed that the Universe 
began in a symmetric state with zero global baryonic charge 
and later (through some baryon-number-violating process, non-equilibrium dynamics, and $\cal{CP}$-violation effects, realizing the three  famous  Sakharov criteria) 
evolved into a state with a net positive baryon number.

As an 
alternative to this scenario, we advocate a model in which 
``baryogenesis'' is actually a charge-separation (rather than charge-generation) process 
in which the global baryon number of the universe remains 
zero at all times.   This  represents the key element of the AQN construction.

In the AQN scenario  the DM density, $\Omega_{\rm DM}$ representing the matter and anti-matter nuggets, and the visible    density, $\Omega_{\rm visible}$, will automatically (irrespective to the axion mass $m_a$ or miss-alignment angle $\theta_0$) assume the  same order of magnitude densities  $\Omega_{\rm DM}\sim \Omega_{\rm visible}$   as they both proportional to one and the same fundamental dimensional parameter of the theory, the $\Lambda_{\rm QCD}$. Therefore, the AQN  model,   by construction, actually resolves two fundamental problems in cosmology  (explains the baryon asymmetry of the Universe, and the presence of   DM with proper density  $\Omega_{\rm DM}\sim \Omega_{\rm visible}$) without necessity  to fit any parameters of the model.

 In other words,  the unobserved antibaryons in visible sector  in this model comprise 
dark matter being in the form of dense nuggets of antiquarks and gluons in the  colour superconducting (CS) phase, which represents a specific type of quark matter (to be distinguished from nuclear matter realized in hadronic phase). 
The result of this ``charge-separation process'' is formation of  two populations of AQN carrying positive (nuggets) and 
negative baryon number (anti-nuggets). We refer to the original papers   \cite{Liang:2016tqc,Ge:2017ttc,Ge:2017idw,Ge:2019voa} devoted to the specific questions  related to the nugget's formation, generation of the baryon asymmetry, and  survival   pattern of the nuggets during the evolution in  early Universe with its unfriendly environment.
We also refer to  very recent lectures  \cite{VanWaerbeke:2026sxs} with a number of applications at different cosmological scales.
%a detail  overview  
%of the relevant AQN physics and typical characteristics of the nuggets such as size, mass distribution, survival pattern, basic constraints, etc.  
 The only comment we would like to make here 
is as follows.  The AQNs are very rare events because they are very heavy in comparison with conventional WIMPs.
The AQN flux     can be estimated  as follows \cite{Lawson:2019cvy}:
 \be
\label{Phi1}
\frac{  d \Phi}{  d A}
=\frac{\Phi}{4\pi R_\oplus^2}  =  4\cdot 10^{-2}\left(\frac{10^{25}}{\langle B\rangle}\right)\left(\frac{\rho_{\rm DM}}{0.3{\rm\,GeV\,cm^{-3}}}\right)
\left(\frac{v_{\rm AQN}}{220~ \rm km ~s^{-1}}\right)\rm \frac{events}{yr\cdot  km^2},
\ee
 where $R_\oplus=6371\,$km is the radius of the Earth, while $\langle B\rangle \approx 10^{25}$ is a typical baryon charge of the AQNs,   and  $\Phi$ is the total hit rate of AQNs on Earth \cite{Lawson:2019cvy}:
\be
\label{Phi}
\Phi
\approx \frac{2\cdot 10^7}{\rm yr}  
 \left(\frac{\rho_{\rm DM}}{0.3{\rm\,GeV\,cm^{-3}}}\right)
\left(\frac{v_{\rm AQN}}{220~ \rm km ~s^{-1}}\right)
\left(\frac{10^{25}}{\langle B\rangle}\right), 
\ee
where $\rho_{\rm DM}$ is the local density of DM within Standard Halo Model (SHM).   
The strongest direct detection limit on the baryon charge $B$ of a nugget is  set by the IceCube Observatory's,  see Appendix A in \cite{Lawson:2019cvy}:
\be
\label{direct}
\la B \ra > 3\cdot 10^{24} ~~~, \la M_{\rm AQN}\ra \approx m_p\la B\ra > 5 g~~~~ [{\rm direct ~ (non)detection ~constraint]}.
\ee
 
 In this work we will use the same typical parameters of the AQN model which had been used in all our previous estimates (including the explanation of  the observed but very  puzzling  UV excess in our galaxy \cite{Sekatchev:2025ixu})  in different environments for dramatically different scales, see review \cite{VanWaerbeke:2026sxs}:
 \be
 \label{eq:values}
   \la B \ra \approx 10^{25}, ~~~  \la M_{\rm AQN}\ra \approx m_p\la B\ra \approx 16 {\rm g}, ~~~ R\approx 2.25\cdot 10^{-5} {\rm cm}, ~~~ \rho_{\rm AQN}=\frac{\la M_{\rm AQN}\ra}{ \frac{4}{3}R^3}     \approx  3.5\cdot 10^{14}\rm \frac{g}{cm^3}. 
 \ee 
 The AQN object could be viewed as a very tiny  neutron star where typical density $ \rho_{\rm AQN}$ of the material is the same nuclear density as in neutron stars. The difference is of course that the neutron star stability   is supported by the gravity, while the stability of the AQN is supported by the axion domain wall with QCD substructure, acting as the squeezer equilibrating the Fermi pressure.

% \subsection{When the AQN hits the Earth}\label{AQN-dense}
For the purpose of this work we need to overview the basic features of the interaction of the AQNs with surrounding material when AQN hits the Earth's atmosphere. We refer to \cite{Zhitnitsky:2025bvy} 
for the introduction into  this topic in context of the BL physics. The only properties we need for the present work are as follow:

i. The annihilation events of the AQN  with surrounding material in atmosphere leads to very high internal temperature of the AQN on the level of $T\approx$ 20 KeV. This implies that the dominant emission from AQNs will be in from of X ray emission, in contrast with  conventional meteorites which assume the temperature in eV range  at most when they propagate in  the atmosphere,   such that they emit photons in optical frequency bands. 

ii. Another  feature   which is relevant for our  present studies is the ionization  properties of the AQN. Indeed, the high internal temperature $T$ of the AQN implies that a large number of weakly bound positrons  
from the electrosphere   get excited and can easily leave the system. As a result, the AQN   acquires a negative  electric  charge $\sim -|e|Q$ with  $Q\sim 10^{11}$. The ratio $eQ/M\sim 10^{-14} e/m_p$ characterizing  this object  is very tiny. However, the charge $Q$ itself is sufficiently large  being capable to   attract the positively charged  ions from air when AQN propagates in atmosphere. 
     
 iii.  One more feature of the AQN propagating in Earth's atmosphere   (which  plays an important role for the present work) is as follows. As we already mentioned   the AQN is absolutely stable object because the energy per unit baryon charge in CS phase (AQN's core) is smaller than for nucleons. In  most  cases an AQN propagates through atmosphere and  Earth's core by  exiting from opposite side of the globe without loosing much baryon charge and momentum as a result of annihilation events. It will be mostly unnoticed as discussed in  \cite{Budker:2020mqk} because the dominant portion of energy released will be in form of the  X rays, axions and neutrinos. However, if such an event is relatively energetic it emits acoustic waves as well. As a result if AQN hits the region with designated infrasound instruments such as 
 Elginfield Infrasound Array (ELFO), all sky camera and a network of seismic stations,  it can be properly  recorded, see details and references in  \cite{Budker:2020mqk}, where such events are identified  with sky-quakes, known for centuries. 
    
iv.   Occasionally, some external strong impact and large energy injection (due to sudden increase of the annihilation rate within the AQN's quark core)  may disintegrate AQN. In this case   a  small chunk from  the original AQN material (in form of the anti-baryon nuclear material) is separated from the original parent AQN, which can be viewed as a separation effect. It will be obviously accompanied with enormous instantaneous energy release (explosion  like). From outside this event looks like a short lasting ``flash".
For the present work precisely this ``flash" plays a key role as this moment of flash will be identified with MT event as discussed in Sect. \ref{sect:proposal}. 
% In a sense, it is very similar to well known and well studied    spallation effect which is very common phenomenon in nuclear physics. 
 The corresponding secondary particles have been coined in  \cite{Zhitnitsky:2025bvy}  as AQN$_s$, where subscript $``s"$ stands for  {\it  s}econdary particle or 
  {\it s}pallation. The secondary particles  have not been invented for the present work with purpose to explain MT events.  
  
  The secondary AQN$_s$  are expected to be much smaller in size than   the parent AQN. The corresponding fragment   really represents  very small chunk of the original (anti-matter) material, but in a different, not CS,  phase
  in comparison with parent AQN.  The difference in binding energies   for parent AQN and daughter AQN$_s$ is the source of energy which should be released at the moment of spallation.   These objects  should have typical sizes $B_{AQN_s}\approx 10^{15}$ which is 10 orders of magnitude smaller than   original parent AQN with typical baryon charge $B\approx 10^{25}$, see (\ref{eq:values}). This typical value $B_{AQN_s}\approx 10^{15}$ is not an ad hoc parameter. Rather it is extracted from  previous BL studies \cite{Zhitnitsky:2025bvy}.  
   
 v.   What is the most important ingredient required for spallation (which will play a key role in this paper  devoted to  MT events) to become very efficient? As argued in    \cite{Zhitnitsky:2025bvy}  the high ionization in the atmosphere is the crucial element for successful  process of spallation.
   This is because   the highly ionized environment dramatically increases the effective strength of interaction of the AQN with surrounding material.
   Indeed,  a high value of electric charge $Q$ attracts more positively charged ions due to large Coulomb cross section. Consequently,  it drastically  increases the internal temperature $T$ leading to 
   further increasing the charge $Q$. This avalanche-like process had been coined in    \cite{Zhitnitsky:2025bvy}  as ``bootstrap" mechanism which becomes effective and operational e.g. during the thunderstorms
   when  the atmosphere in thunderclouds is known to become highly  ionized. As we discuss in next Sect. \ref{BL-features} this is precisely the reason why BL events are strongly correlated with thunderstorms.

   \section{The AQN framework: application to   the BL events}\label{BL-features}  
       As we already mentioned the BL events are close cousins of the MT events. Therefore, it is quite natural to explain the basic  features of the BL events as observed for centuries within AQN framework before moving to formulation of our proposal in Sect. \ref{sect:proposal} on the nature of MT events. 

  The BL phenomenon  has been known for centuries,  see recent book \cite{Herbert_Boerner} and review papers \cite{SMIRNOV1993151,SHMATOV2019105115,Rakov_Uman_2003,hgss-12-43-2021} with large number of references therein.  
 There are many mysterious properties  of the observations  \cite{Herbert_Boerner,SMIRNOV1993151,SHMATOV2019105115,Rakov_Uman_2003,hgss-12-43-2021}
 which are impossible  to understand if interpreted in terms of the conventional    physics. 
 In fact a complete failure to understand even the very basic features of the BL phenomenology (such as required power, or passing through a solid glass) enforced the  researchers to look for   possible answers to subatomic physics, well outside the conventional mechanisms considered in the past. In particular,  in ref. \cite{Stephan:2024mau} it was   suggested that the magnetic monopole might be powering the BL, while in ref.  \cite{Ralston:2024xlu} this idea was modified  by adding an electrical  charge into the system by making  the  dyon (magnetic monopole with non-vanishing electric charge).  Our  proposal  \cite{Zhitnitsky:2025bvy} is also deeply rooted into subatomic physics, but in dramatically different way, as  briefly reviewed  below.

  The proposal   \cite{Zhitnitsky:2025bvy}  has been    formulated as follows. The secondary particles (after   spallation) in form of the  antimatter AQN$_s$ are  identified with Ball Lightning events, i.e.
    \be
    \label{eq:proposal}
  \rm   secondary ~AQN_s ~events ~~~\equiv ~~~ Ball~ Lightning ~events. 
     \ee
The main goal of this  section    is to overview   the various BL  observations    and show how they   could  be  naturally   explained      if one accepts the proposal
 (\ref{eq:proposal}).   
 
 \subsection{Association of the BL events with thunderstorms}\label{sect:thunder}
 Association of the BL events with thunderstorms is a very natural outcome of the process of spallation (flash) as described in items iv. and v. at the end of Sect. \ref{AQN}.
 The main ingredient for  spallation (flash) to become efficient is sufficiently    high ionization of the atmosphere in the region where AQN hits. Only in this case the avalanche-like process becomes operational and spallation occurs. 
 The ionization is indeed becomes very high under the thunderclouds which is a natural  explanation for  correlation between BL events with thunderstorms. Only in the case when a small secondary piece coined as  $AQN_s$ starts to propagate in atmosphere it may  slow down to small velocities $\sim\rm  m/s$ from enormous $\sim 200~\rm  km/s$ which  represents original typical DM velocity. This journey is 
 sufficiently short    (few km scale)  when BL reaches the Earth's ground   and emits the visible light such that it  can be  observed by eyewitnesses. 

 If spallation does not occur, and AQN$_s$ are not formed, the parent AQN continues to propagate with (almost) original DM velocities $\sim 200 \rm km/s$ and  cannot be observed  in visible frequency bands.  The corresponding objects have been identified with sky-quakes  and could be studied    in  infrasound frequency bands  as discussed in  \cite{Budker:2020mqk} and briefly reviewed  in \ref{sect:cousins}. 
 
 One should emphasize that the BL events are commonly  (incorrectly) assumed to be   originated  from thundercloud's plasma  (formed during a thunderstorm).  This assumption is based on observed correlation between BL events and thunderstorms. Our interpretation of the same correlation is dramatically different: it emerges  due to the high ionization of the air in thunderclouds when a propagating AQN experiences the spallation process
 forming the secondary AQN$_s$.  In particular, the BL events could emerge under clear sky if ionization of the air is sufficiently high due to some different reasons. This comment is crucial in our discussions  in Sect. \ref{sect:5} of the AQN induced events and their relations to   subclass of the UAP events  because some UAP events were thought could  be the BL events. However,  such identification was generally dismissed   due to conventional (but incorrect) argument that BL may appear exclusively during the thunderstorm and cannot emerge under the clear skies.

 \subsection{Source of the energy powering BL}\label{sect: energy}
 The source of the energy in the AQN framework is obviously the antimatter annihilation with surrounding material. 
   The mean energy for BL is estimated in  \cite{SMIRNOV1993151} as $2\cdot 10^2 \rm kJ$,  
   which corresponds to baryon charges $B_{\rm AQN_s}\approx 10^{15}$ of small chunks of antimatter after spallation, i.e.  
      \be
   \label{s-energy}
 E_{\rm BL}\sim  B_{\rm AQN_s} (2 m_p) \approx 3 \cdot 10^2 \rm   kJ, ~~~~~ J= 0.6\cdot 10^{10} \rm GeV. 
\ee
 This represents total  BL energy   emitting   in all frequency bands during its life time $\tau\approx 10$s.
 However, BL will be mostly emitting the X ray photons due to its high $T$ in keV range.  
 
   \subsection{Size of   BL(in visible frequency bands).}\label{sect:radiation}
   The X ray photons emitted by AQN$_s$ will be absorbed by $O, N$ atoms (atomic photoelectric effect) in atmosphere. The process will be accompanied by electron emission, which will be quickly absorbed by atoms in air on very short distances.  These excited and ionized atoms and molecules (made of  Oxygen and Nitrogen) will emit visible light which is observed as radiation coming from BL according to our proposal (\ref{eq:proposal}). The visible size of BL is, therefore,  determined by the mean-free path of the X rays in atmosphere as estimated in \cite{Zhitnitsky:2025bvy}:
      \be
   \label{lambda1}
   \lambda^{O,N} \rm \approx 10 ~  cm,   ~~~~~~~~~~~~~~ [ to~ be~ identified~ with~visible~ size ~of~ BL]. 
   \ee
    This picture of emission  is consistent with observation  that BL emits UV and  x ray radiation along with visible light as discussed above. In fact, UV and x rays are originated from the the core of the $AQN_s$, in contrast with visible light 
   which is a secondary process in the AQN framework as described above.  This  picture is perfectly consistent with    the presence of UV or x rays   which had been directly observed \cite{STEPHAN201632}.
  From processes described above it is obvious that the power in visible frequency bands represent a very small portion (on the level $10^{-3}$ or even less) of the  total power of the BL's emission.

      \subsection{BL passing through glass windows. The BL's   new scale of the problem.}
      \label{sect:glass}
      One of the most mysterious property  of the BL is its observed passage through glass windows. In scientific literature it had been recorded by 
       the authors of ref. \cite{BYCHKOV201669}.   Using the modern instruments (such as  optical and  scanning microscopes and laser beam probing the glass) the authors   have found the traces which  are  left by 20 cm BL passing through the window glass.  The authors  discovered a cavity of 0.24 mm  diameter, see Fig.3 in that paper.  
      The authors  correctly interpreted this event as an undeniable   evidence of  a ``material" nature of BL. The scale of this 
      cavity is dramatically different from the  20 cm scale   of the BL  as observed at visible frequency bands. 
      
      In the AQN framework the emergence of this new scale (0.24 mm in comparison with 20 cm scale as  observed in visible light) has a very natural explanation. Indeed, the mean free path  of the X rays in glass is:
       \be
   \label{lambda2}
     \lambda^{Si} \rm   \approx 0.25 ~  mm, ~~~~~~~~~~~~~~ [ to~ be~ identified~ with~ size ~of~a~ cavity~ in ~ glass],
   \ee
   which is very close to observed cavity size.  The energy density injected at the instant when  BL passing through the window glass of width $  l\approx \rm 2mm$ can be estimated from (\ref{lambda2}) as 
   $ \epsilon^{Si}\approx 10^3 {\rm kJ}/{\rm cm^3}$ which is 
 more than sufficient to melt the glass in small volume to allow BL to pass through glass windows.

\subsection{Frequency of appearance} \label{sec:event_rate} 
The main question which is normally being   asked  regarding the BL rate  is as follows: 
  It is known that BL are associated with thunderstorms.  Therefore, a conventional (but very naive) argument  would suggest that the lightning itself should play an important role in formation of the BL.  Then,  why the  observed frequency  of BL events is much  lower than an average frequency of conventional lightnings in the continental U. S., which is   about $24~ \rm km^{-2}~yr^{-1}$? We rephrase the same question in a different way: what is so special about very rare lightning events which produce BL   in comparison with vast majority of lightning  events which do not lead to BL? Furthermore, the observed BL events are not correlated with the strongest and most powerful  lightning events. 
  
 In the AQN framework  the answer to this question lies in our understanding of BL events as AQN-induced events. To estimate the frequency of the BL events we should multiply flux  (\ref{Phi1}) to  parameter ${\cal{F}}$ which describes the fraction of time when the area $\rm d A$ has been under
   thunderclouds (when the ionization is high), i.e. 
  \be
  \label{BL-flux1}
\frac{ d \Phi^{AQN}_{BL}}{  d A} 
 \approx  4\cdot 10^{-4}    \rm \frac{(BL~events)}{yr\cdot  km^2}, ~~~~  {\cal{F}}\approx 10^{-2}
\ee
where estimation for parameter   $ {\cal{F}}\approx 10^{-2}$ 
 is based 
on compilation of the annual thunderstorm duration from 450 air weather system in USA as described  in 
 \cite{Gurevich:2004km}.   The corresponding estimates suggest that on average the thunderstorms last about $1\%$ of  time of the year in each given area  \cite{Gurevich:2004km}.  
 In this estimate we also assume that every thunderstorm produces sufficiently high ionization in the area such that spallation becomes 100  \% efficient and operational, which is likely to be an overestimation.

We consider our estimate (\ref{BL-flux1}), to be treated as an upper bound,    is    perfectly consistent with lower bound  as given in  \cite{Stephan:2024mau,STEPHAN2022105953}.
In fact, it is amazing that so different estimates which include dramatically different environments and physical systems 
(from thunderstorms and lightning events to dark matter density $\rho_{\rm DM}$ and galactic wind velocities explicitly entering all  the estimates) are consistent with each other.

  Precisely the estimate (\ref{BL-flux1})   answers (within the AQN framework) the question formulated in the first paragraph of this subsection: why the BL events  are so rare? The proposed  answer \cite{Zhitnitsky:2025bvy} is that  the rareness of  BL   events is a consequence of the rareness of the  AQN events with  very tiny DM  flux (\ref{Phi1}).

         \section{Proposal (\ref{eq:proposal-MT}) confronts  the observations}\label{sect:proposal}
         This is the main section of the present work. We identify the moment of spallation  which manifests itself as an instant of explosion (instant of a short lasting ``flash") with  large amount of energy (to be estimated below)   will be released into space, i.e.
 \be
    \label{eq:proposal-MT}
  \rm    spallation    ~of~AQN~ (short ~lasting ~``flash")  ~~~\equiv ~~~ Mysterious ~ Transient ~event. 
     \ee 
     In what follows we shall argue that all items 1.-5. from Sect.\ref{items} find their natural explanation if the proposal (\ref{eq:proposal-MT}) on nature MT events is accepted\footnote{We should mention here that the results \cite{2022MNRAS.515.1380S,Villarroel_2025,Bruehl:2025} are still a matter of debates as this analysis has been criticized  by \cite{2026arXiv260121946A} pointing to some deficiencies in dataset selections. This criticism   was dismissed in \cite{2026arXiv260215171V,2026arXiv260418799B}. We do not wish to be a part of these debates. Rather we want to emphasize  that the basic elements of our identification (\ref{eq:proposal-MT}) within AQN framework such as spallation (flash) event and its characteristics, its sensitivity to ionization of the environment had been previously discussed in context of the BL phenomena as reviewed in Sect.\ref{BL-features}. The   qualitative order of magnitude estimates in this work do not depend on statistical subtleties of the debates \cite{2026arXiv260121946A,2026arXiv260215171V,2026arXiv260418799B}. Therefore, we opted to keep all our numerical estimates below using the results of  the original papers \cite{2022MNRAS.515.1380S,Villarroel:2019bky,Villarroel:2020knw,Villarroel:2021ddh,2024MNRAS.527.6312S,Villarroel_2025} for consistency of presentation.}. We follow exactly the same order of Sect.\ref{items}, item by item, which will be numerated below as subsections 4A-4E.  We also explain the relation between MT events, BL events, pseudo-meteorite events,  and sky-quakes. All these puzzling and mysterious events are in fact closely related with each other, and the rareness of the corresponding events 
   is proportional to one and the same     very tiny DM  flux (\ref{Phi1}) within AQN framework, similar to estimates for BL rate as given in Sect. \ref{sec:event_rate} above.   
     
     \subsection{Nature and Source of Energy powering  MT events}\label{sect:1}
    An  AQN enters the Earth's atmosphere with a typical for DM velocity $v_{\rm AQN}\sim 10^{-3} c$. In majority of cases (as discussed in item iii at the end of  Sect.\ref{AQN}) it propagates through atmosphere and Earth's core without loosing much of momentum and baryon charge
     as the cross section of the interaction  is relatively small, and  can be well approximated by its geometrical size $\pi R^2$, where $R\approx 2.25\cdot 10^{-5} \rm cm$ is the AQN's core size   for a typical $B\approx 10^{25}$ according to (\ref{eq:values}). 
     Such events can be hardly noticed by eyewitnesses as this  object does not emit  much  in  visible frequency bands. Instead, it mostly radiates  the X rays \cite{Budker:2020mqk}.  Nevertheless, in rare cases when the AQN  is sufficiently large in size with $B\approx 10^{28}$ such  events can be observed due to their strong shock  waves which can be recorded by dedicated  infrasound instruments such as Elginfield Infrasound Array (ELFO). Such powerful AQN events had been identified with sky-quakes which had been known for centuries without given any physics explanations, see Sect. \ref{sect:cousins} for references and details, including the description of the puzzling event recorded by ELFO  on July 31-st 2008 in vicinity of London, Ontario, Canada.
     
     However, in  rare cases when a typical AQN enters the region with high level of   ionized air  the spallation effect as described  by items iv and v at the end of  Sect.\ref{AQN}, may occur. We expect this could happen at relatively low altitude, 10-30 km because the air density should be   sufficiently high for spallation mechanism (avalanche type instability) to be efficient. In this case AQN disintegrates  to two (or more) pieces, the parent AQN and the secondary AQN$_s$ with $B_{AQN_s}\approx 10^{15}$. These secondary 
     AQN$_s$ will slow down very quickly, in contrast with large parent AQN,   and can reach the ground region from location of spallation in a matter of seconds. These secondary AQN$_s$, when they reach the ground, are identified with the BL events according to 
  (\ref{eq:proposal}). All observed characteristics   of the BL events are consistent with this proposal as briefly reviewed in Sect. \ref{BL-features}. The most important feature now is that a strong correlation between BL events and thunderstorms finds its natural explanation in this framework as the air during a thunderstorm  is known to be   highly  ionized. Therefore, the spallation is likely to occur, and  secondary AQN$_s$ is likely to be formed.  The typical baryon charge  of the secondary AQN$_s$    was estimated from total energy  of BL (\ref{s-energy}) which corresponds to   $B_{AQN_s}\approx 10^{15}$ which will be  our benchmark for MT events. In other words, we use the same size of the secondary AQN$_s$ extracted from BL events  in our estimates below for analysis of the MT events. 
  
  Now we   estimate the total energy being released at the moment (instant) of spallation which we identify with MT instantaneous event according to (\ref{eq:proposal-MT}). 
    \be
   \label{MT-energy}
 E_{\rm MT}^{\rm total}\sim  B_{\rm AQN_s} (\Delta E) \approx  (10^2-10^3) {\rm   J}, ~~~~  \rm  \Delta E \approx (E_{quark}-E_{nuclear})\approx (1-10) MeV, 
\ee
     where   the secondary small piece AQN$_s$ assumes    a nuclear matter phase (not   CS quark matter phase) as explained in   \cite{Zhitnitsky:2025bvy}. 
     In estimate (\ref{MT-energy}) we assume that the difference in energy scales (which includes many different components contributing to the energy such as chemical potential, binding energy, domain wall pressure, etc) between two different phases should be in $(1-10) $ MeV range which represents a typical scale in nuclear physics.   
     The total energy (\ref{MT-energy})  should be compared with observed energy in optical bands. For $m_{AB}\in (17-19)$ at distance $d\approx 10 $ km the luminosity in optical bands is estimated to be around $L_{\rm optical}\approx 10^{-6} W$. For larger $d$ it scales correspondingly $\propto (d/10 ~\rm km)^2$. Therefore the   energy ${\rm E}_{\rm MT}^{\rm opt}$ and number of instantaneously emitted photons $N_{\gamma}^{\rm opt}$ at the moment of  MT event in optical frequency bands from observations within AQN framework can be estimated as 
            \be
   \label{MT-optical}
 {\rm E}_{\rm MT}^{\rm opt}\approx   L_{\rm opt}\cdot  \left(\frac{d}{10\rm km} \right)^2\cdot 30 {\rm min}\approx 2\cdot 10^{-3} J,~~~~~ N_{\gamma}^{\rm opt}\approx\frac{\rm  E_{\rm MT}^{\rm opt}}{3\cdot 10^{-19} J}\approx 10^{16}  \left(\frac{d}{10\rm km} \right)^2\rm    .
\ee
where 30 min is a typical time of exposure recorded by POSS-1. 
   This estimate  implies that only very small portion ($  10^{-5}$ or less) of the MT total energy (\ref{MT-energy})  is released in form of the visible light. The dominant portion of the energy will be released in form of the axions, low energy neutrinos, X-rays and heat.  One should also  comment  here that a similar  argument applies to  BL events within the same AQN framework  
     as reviewed in Sect. \ref{BL-features}. This small factor between optical and x ray emissions can be understood (in both cases, MT as well as BL events) as a result of series of chain processes when emission of a single  X ray photon (with energy $\sim 10$  keV) by AQN generates the emission of a single optical photon (with energy $\sim 1 $ eV). These visible photons  can be  recorded by POSS-1. The estimate of  visible size of the MT events  in all respects is  similar to estimation of the BL events,  as briefly reviewed in Sect. \ref{sect:radiation}.

     \subsection{MT events. The spatial and temporal  scales}\label{sect:2}
     The process of spallation is very short on the level of a fraction of micro-second or even less, such that it would look like a point-like explosion (flash) from  a telescope even from relatively short distance 10 km or so. Indeed, the longest time scale in a series of of processes leading to the emission of the optical photon as mentioned at the end of   previous section  \ref{sect:1} is the life time of a free electron emitted as a result of photo-effect.  This time scale  in air  is  very short, around $\tau\approx 0.1\mu \rm s$, see e.g. \cite{Gurevich_2001} such that free electron will be quickly absorbed by atoms in air on very short distances around $10^{-2} \rm cm$ much shorter than mean free path for X rays as given by (\ref{lambda1}). These excited atoms and molecules will consequently emit the light in visible frequency bands, which  can be recorded by POSS-1.   Therefore, if an AQN moves with typical for DM velocity $\sim 200 \rm km/s$ the displacement (leading to asymmetry) of the source AQN$_s$ during this short period of time is sufficiently     small, i.e.
     \be
    \label{MT-displacement}
      \Delta L\sim V_{\rm AQN}\cdot \tau \sim 2 \rm cm, ~~~~~~~ 1" (arcsec) =4.8 \left(\frac{d}{10\rm km} \right) cm, 
     \ee
      where for reference we also express arcsec in terms of the physical distance scale  as SuperCosmos resolution employed in \cite{Villarroel_2025} is $0.67$ arcsec/pixel. Comparison of the physical scales  (\ref{MT-displacement})  shows   that the observed image of the MT event recorded by POSS-1 will be mostly spherical and point-like (with possible mild elongation of order $\Delta L$) as the observed size of MT event  in optical bands   is determined by mean free path of X rays in atmosphere similar to analysis of the BL events (\ref{lambda1}). 
      
 Our next comment is related to observed      full-width-half-maximum (FWHM) for  MT events as recorded by POSS-1.  It varies from 2.8-8 arcsec \cite{Villarroel_2025},
 which is consistent with our interpretation of the observed image of the MT in optical bands as mean free path of soft X rays in atmosphere (as a result of spallation), which 
 assumes $\sim 10$ cm scale\footnote{\label{footnote:size}The observed visible size of the object dramatically depends on spectrum  of the  X rays and altitude where spallation occurs. The X ray spectrum as a result of spallation is expected to be very broad, from few keV to $10^2$ keV.  The visible size of the object is obviously determined by the lowest x ray frequency as it dominates the x ray absorption because the atomic photo-effect cross section scales as $\omega^{-7/2}$, see \cite{Zhitnitsky:2025bvy} in the context of BL phenomena. The excited and ionized atoms and molecules will emit the visible light (as observed by POSS-1) at much shorter time scales,
 such that mean free path of  X rays determines the visible size of the objects similar to our studies of the BL events as reviewed in Sect. \ref{sect:radiation}.   In particular, the mean free path for 2 keV X ray at altitude 10 km is $\lambda \rm (2 keV, 10km)\sim 4.6~ cm$, while 
 $\lambda \rm ( 3 keV, 10 km)\sim 20 ~cm$.  This spectrum is drastically  different from the x ray spectrum of the AQN$_s$  in steady state (slow motion) when the spectrum is actually determined by the internal temperature of the nuggets, similar to our estimate of the BL size (\ref{lambda1}). Altitude also plays a role in estimate of the visible size of the object as the density of air decreases with altitude, while mean-free path increases correspondingly.} as observed from   distance $d\approx (10-30) \rm km$. The measurements of the FWHM values for MT events also suggest that they must be sub-second flashes of light \cite{Villarroel:2025xym,2026arXiv260320407B}. This is obviously consistent with the proposal (\ref{eq:proposal-MT}) when MT events are characterized by very short time scales on the level of $\mu$s or even less.

      To summarize this subsection: the proposal (\ref{eq:proposal-MT}) is consistent with observed   spatial and temporal  scales for MT events within the AQN framework. 
      The MT events as observed in optical bands are mostly spherically symmetric (with possible mild elongation of order $\Delta L$) with size $(10-10^2)$ cm which is mostly determined by frequency  of emitted X rays and the  altitude where  the MT event occurs, see footnote \ref{footnote:size} with details. Very short life time of spallation process (instantaneous flash) is also consistent with observed 
      FWHM measurement which is sharper and more circular than the typical images of stars.
      
      \subsection{MT events:  alignment    within a single exposure}\label{sect:3} 
     It has been claimed that in a number of cases the multiple MT events   are well aligned  along a line \cite{Villarroel:2021ddh,Villarroel_2025,Villarroel:2025xym}.
     The probability of two such MT events (assuming random distribution)  appearing in the same $(10\rm ~ arcmin)^2$ box is $p\sim 10^{-6}$   \cite{Villarroel_2025}.  The  probability  for aligned events within a single exposure should be even much  smaller if MT  were random events. The observed alignments are in fact very precise: a 3-point alignments shown in \cite{Villarroel_2025} are aligned with 1-2 arcsec, while 5-point alignments are aligned within 10-15 arcsec. What could be the nature for observed multiple MT events and their alignments within AQN framework if proposal (\ref{eq:proposal-MT}) is adopted?
     
The answer lies in the nature of spallation of the parent AQN with $B\sim 10^{25}$ to two pieces: the parent AQN and   the secondary AQN$_s$ which has  much smaller $B\sim 10^{15}$. It is quiet obvious that if appropriate conditions occur for spallation (high ionization of the air),  as described in items iv and v at the end of Sect. \ref{AQN}, a first spallation can be followed by consequent   event which may occur  within the same region. The spallation events could  occur multiple times from the same parent AQN.  One should mentioned here that multiple BL events are known to occur, and had been described in the literature as  reported by eyewitnesses.  From AQN perspective the presence of multiple MT events  is very natural as well as multiple BL events is very natural outcome of the proposed mechanism. 
%In both cases  corresponding to multiple 
%secondary AQN$_s$ which end up with multiple BL events when secondary AQN$_s$ slow down, reach the ground and could be observed by eyewitnesses demonstrating a number of highly nontrivial properties. These  features  had been collected for years (centuries) as 
%reported in  \cite{Herbert_Boerner,SMIRNOV1993151,SHMATOV2019105115,Rakov_Uman_2003,hgss-12-43-2021}. All these properties could be understood simultaneously with AQN framework  as explained in \cite{Zhitnitsky:2025bvy}. 

Furthermore, a very massive parent AQN does not change much momentum after spallation, and its motion continues along the same trajectory, which we identify with a line of alignment  of multiple MT events. The deviation from exact line is mostly determined by location of affected area  at the instant of spallation.  If    the secondary AQN$_s$ were located exactly  at    front of the parent AQN (with respect to direction of the motion) it would be perfect alignment as small piece AQN$_s$ will move along the same direction   as parent AQN. However, if the secondary AQN$_s$ were  located slightly off centre, it could assume a small momentum in a transverse  direction, and can be recorded as an independent MT event. 

To summarize: the presence of multiple MT events obviously implies Non- Poissonian distribution. It    manifests itself in presence of clustering events   describing  multiple MT events as reported in \cite{Villarroel:2021ddh,Villarroel_2025,Villarroel:2025xym}. We presented the arguments that the presence of such clusters is a very natural outcome of  the proposal (\ref{eq:proposal-MT}). Furthermore, our arguments suggest that the majority of all multiple events appearing in the same $(10\rm ~ arcmin)^2$ box are related to the   clusters. Each cluster should be thought as a set of multiple spallation events induced by one and the same parent    AQN, which are observed as MT events.   Therefore, these multiple events by no means can be treated as random events. Instead, they should be treated as the clusters when Non- Poissonian distribution must apply. 

\subsection{MT events: strong correlation with nuclear testing}\label{sect:4}
Another puzzling feature  which was discovered in \cite{Bruehl:2025} and independently confirmed in \cite{2026arXiv260400056D} is that MT events are strongly correlated  with nuclear testing.  This is absolutely amazing finding which is extremely hard to understand within conventional physics \cite{Bruehl:2025}. What is even more puzzling  is that the correlations with MT events are observed for the day of testing and day after (not before) the testing, which explicitly shows  the causality direction of these (naively)  unrelated  events. 

While the observed correlation looks very puzzling within conventional physics, it is a very natural outcome within AQN framework if proposal (\ref{eq:proposal-MT}) is adopted. Indeed, the nuclear testing lead to very high level of  ionization in atmosphere as a result of very large number of processes, see details below. At the same time high level of  ionization in atmosphere will dramatically increase the likelihood of the spallation as explained in items iv and v at the end of Sect \ref{AQN}. The well established correlation between the thunderstorms (which are characterized by high level of ionization) and BL events is the direct consequence of the same phenomena  when the secondary AQN$_s$ are formed as a result of spallation. They  slow down, reach the ground and eventually manifest themselves as BL events as described in  \cite{Zhitnitsky:2025bvy} and briefly overviewed in Sect. \ref{BL-features}.  

The nuclear tests are known to ionize atmosphere, and it is well recorded in the literature, see classical textbooks   \cite{glasstone1962effects,osti_6852629} on the subject.
The most important  mechanism for atmosphere's ionization which may  last  sufficiently long time, from hours to several days (the time scale relevant for these studies),    is related to fission products as a result of nuclear test. In this case the  produced unstable isotops of Kr, Xe, I and dozens  of other elements undergo beta decays, emitting energetic electrons. A beta electron ionizes   of thousands of molecules along its path before stopping.  There are many other sources of ionization which may contribute   to ionization of  the atmosphere which may last for days, such as neutron activation and secondary gamma particles. 

To summarize: the ionization of the atmosphere after nuclear testing is well established phenomenon, similar to high ionization of the atmosphere during the thunderstorms. In both cases it leads to dramatic increase of the likelihood of the spallation, which is identified with MT event  according to  proposal (\ref{eq:proposal-MT}). We estimate the MT event rate based on this observation in Sect. \ref{sec:MT_event_rate}.
 
      \subsection{MT events: strong correlation with UAP reports}\label{sect:5}
     The correlation between nuclear tests and UAP reports had been recorded for decades, see comprehensive recent review \cite{knuth2025newscienceunidentifiedaerospaceundersea}. So, it is not really a surprise that the MT-UAP  correlations had been found in  \cite{Bruehl:2025} because the correlations between MT events and nuclear tests had been  discovered and confirmed as discussed in previous Sect.\ref{sect:4}. Nevertheless, we opted to elaborate on this 
     matter because it provides one additional argument supporting the proposal (\ref{eq:proposal-MT}) as it identifies a specific type of UAP events which could be directly 
     linked to AQN$_s$ spallation events, and which according to the proposal (\ref{eq:proposal-MT}) are identified with MT events. Furthermore, the elaboration on MT-UAP correlation may play an important  role in design of specific instruments which could play an important role for systematic studies of the UAP events as recently discussed in \cite{doi:10.1142/S2251171723400068} in context of the Galileo Project or \cite{Szydagis:2024eea} in context of UAPx project. This elaboration also sheds some light on close cousins of the MT and BL events which represent specific class of 
       the  UAP events as discussed in Sect.\ref{sect:cousins}. 
     
       \subsubsection{theoretical expectations}\label{theory}
    From AQN perspective the secondary AQN$_s$  when it propagates in atmosphere could be classified  as a specific class  of the UAP events. If the secondary AQN$_s$ has sufficiently small baryon charge (mass)
     of order $B_{\rm AQN_s}\lesssim  10^{15}$ it could   stop before reaching the ground as a result of friction and annihilation processes along its path. These events could stop at altitude of few kilometres or close to the ground where could be  observed. In the last case these events had been identified with BL events as proposed in \cite{Zhitnitsky:2025bvy} and briefly overviewed in Sect. \ref{BL-features}. In the former case they also behave as slow moving BL objects, however from a distance of few  kilometres the BL object could look like a stationary bright object. 
     
     If baryon charge  is few orders of magnitude larger, let us say
     $B_{\rm AQN_s}\gtrsim  10^{18}$ the corresponding object cannot stop before it hits the ground  as there is no enough material along the way to complete annihilations in atmosphere. As a result
   AQN$_s$ continues to move with relatively high velocity at low altitudes. The visible size of this moving object is determined by the mean-free path of the x rays, similar to BL events as discussed in \cite{Zhitnitsky:2025bvy} and reviewed  in Sect. \ref{sect:radiation}.  When this object hits the ground (solid soil or water), the antimatter material will heat (and melt) the   surrounding region of impact on the ground as a result of annihilation processes. There should be no any remnants  of material  in the affected area due to   annihilation processes, in contrast with meteorite events. Only the heat, excess of radiation and strongly ionized air will remain in the area for some period of time. Strong electric and magnetic disturbances could be also observed due to large electric charge carried by AQN$_s$, similar to BL events as discussed in  \cite{Zhitnitsky:2025bvy}.
   
    If AQN hits the  water's surface it will continue to propagate with the same initial velocity as the core size of the nugget (not to be confused with   observed size of the object in visible frequency bands) is so tiny according to (\ref{eq:values}),  that there will be no any waves which would normally occur for such air-water crossing event.   It continues to move downward   to reach  deep underground region where   complete  annihilation eventually occurs.    In what follows we focus precisely on this specific type of UAP, and its observable manifestations. There are many other UAP events which may have very different characteristics, and which require  
   different type of instruments to study them, see e.g. the Galileo Project \cite{doi:10.1142/S2251171723400068} or UAPx project \cite{Szydagis:2024eea}.
   %These UAP events may or may not be close cousins of the MT events as discussed in Sect.\ref{sect:cousins}.

      \subsubsection{some observations}\label{observations}
     We start by quoting several paragraphs from Section 3.3.5 of \cite{knuth2025newscienceunidentifiedaerospaceundersea} titled ``Green Fireballs and Project Twinkle",
     with description of observations  which closely resemble the manifestations    of the AQN$_s$ events as given in previous Sect.\ref{theory}.      
 In late November of 1948 a   number of green fireballs had been reported, see large number of references  in  \cite{knuth2025newscienceunidentifiedaerospaceundersea}.
 The investigation started when  Kirtland Intelligence Officers contacted astronomer and meteor pioneer, Lincoln La Paz
from the University of New Mexico. Now we quote few paragraphs from Section 3.3.5 of \cite{knuth2025newscienceunidentifiedaerospaceundersea} to give a flavour of the observed events given by professionals: ``La Paz insisted that the
green fireballs could not be meteors because their trajectories were too flat,
their colour too green, and that no meteoritic material had been recovered".  Another quote is: ``But the biggest mystery of all was the fact that no particles of a
green fireball had ever been found. If they were meteorites, La
Paz was positive that he would have found one."  Another unusual factor was a large size of the green fireballs which was described as ``terrifying".  An enormous observed velocity 
of the  the green fireballs, which is not typical for the meteorites, was also emphasized by eyewitnesses. 
Such large meteorites moving with enormous velocity normally are accompanied by sound and shock waves that
break windows. Yet,  the observers reported that they did
not hear any sound.
One additional quote:
``if the green fireballs were meteors, then they should have been observed
globally --certainly not limited to a handful of sensitive nuclear sites in New
Mexico". 
    
    We conclude this subsection with description of     final moment of another    fire-ball  event when it hits the ground and melts material in surrounding area.  The event occur  in US town of Elma (Washington State) on July 15, 2003 as described in \cite{2020arXiv201200686O}. 
    The key elements of the Elma event are as follow  \cite{2020arXiv201200686O}: 1.after the event (in form of a bright fireball) the   eyewitnesses follow the trajectory of the fireball to find a spot where it hits  the ground. The observers found the  glassy black rocks on the spot; 2. the  rocks were very hot (in fact, one of the witnesses had burned a thumb and a finger by collecting the rocks); 3.  the discovered stones had been analyzed by the University of Washington. The conclusion was     that the glassy rocks were not meteorites.  Instead, the chemical composition of the rocks  was very similar to soil of surrounding area\footnote{\label{footnote:meteorites}There are many reports in the literature describing the meteor-like events which  (after detail studies by professionals) turned out to be not the meteorites. The corresponding events are classified as ``Pseudo-meteorites  events", see   \cite{2020arXiv201200686O} with large number of  references and details.}. 
  
  \subsubsection{theoretical prediction confronts the observations}\label{comparison}
  We now  compare the theoretical expectations  as  presented in  section \ref{theory} with small sample of observed events recorded by professionals as briefly described in section \ref{observations}. Our conclusion is that  we find full qualitative consistency   between theory and observations. Indeed, the observed size of the objects in visible frequency bands is determined by the mean free path of X rays, see footnote \ref{footnote:size} with few comments. It could easily be very large, 1-10 meters and even larger\footnote{\label{footnote:size1}One should contrast  the  estimate of the mean free path for the x rays as a result of spallation (explosion like) as explained in footnote \ref{footnote:size} 
  from   the estimate of the x ray mean free path for AQN$_s$ in steady state, similar to slow moving BL. In this case a typical internal temperature of the AQN$_s$ could reach 20 keV, see item i. at the end of  Sect.\ref{AQN}, 
  such that the typical x ray frequency is also around 20 keV. The mean free path for such energetic photons at 10 km altitude is enormous, $\lambda( \rm 20 keV, 10 km)\sim 30 m$. Precisely this numerical estimate represents  an apparent  visible size of the  AQN$_s$ moving downward after spallation, and identified with ``green fireballs" from the text.}.  This scale determines the visible size of the object because the x rays after a chain of processes produce   visible photons which can be observed by eyewitnesses. This size  has no relation to actual core size  of the $AQN_s$ which is much smaller of order $\mu m$ as estimated in  \cite{Zhitnitsky:2025bvy} in context of the BL phenomena. The ``terrifying" size of the objects as described by eyewitnesses reflects the   size of the $AQN_s$ in visible frequency bands\footnote{\label{footnote;spectrum}While the estimation of an apparent  size of moving AQN$_s$ is straightforward, the  precise computations of the spectrum (including the green colour of the object) of this emission is much more  complicated problem due to a large number of competitive processes: direct ionization by moving negatively charged AQN$_s$ (along with X ray ionization), attraction of positively charged ions by moving $AQN_s$, exchange reactions which excite ionized atoms and molecules with consequent emission of the  UV and visible light. These processes strongly depend on the velocity of the moving AQN$_s$, its charge, state of atmosphere, altitude, etc. In particular, for high velocity of the  AQN$_s$ the dominant emission is likely to be in UV, while for lower  velocities it is likely to be in visible frequency bands.  All these questions are well beyond  the scope of the present work.}.  
  
 The observed  large size of moving $AQN_s$  as explained above  are not related to any solid substance, in contrast with conventional meteorites which could weight many tons for meter-sized objects.   As a result,  fast  moving AQN$_s$ with tiny weight  is not capable to produce an observable sound waves. We should comment here that a much larger AQN may indeed generate acoustic  wave with low intensity and in infrasound frequency bands as  mentioned in item iii.  at the end of Sect. \ref{AQN}.
       
   The very high velocity (in comparison with meteorites) of the observed fireballs  is also easy to explain in our framework: an initial velocity of the AQN when it enters the atmosphere is $\sim 10^{-3}c$, which is  at least one order of magnitude larger than conventional  velocity of a meteorite.  Even when $AQN_s$  slows down in lower altitudes\footnote{This motion is characterized by enormous deceleration easily exceeding $ 10^2 \rm g$ as estimated in  \cite{Zhitnitsky:2025bvy} in context of the BL physics.}, the velocity  remains to be much higher  in comparison  with conventional meteorites  as a result of   high initial velocity. 
      
      The collected  hot melted rocks at the spot as described in Sect.\ref{observations} has  also very natural explanation within our proposal. Indeed, when $AQN_s$ hits  
      the soil its effective size dramatically decreases (and temperature increases) together with its mean free path in rocks as the density of material increases by 3 orders of magnitude, in comparison with air. The internal temperature greatly increases such the surrounding material can be easily melt. 
      The excess of radiation should also stay in the area of impact.  In all  respects the heating of the surrounding material and excess of the radioactivity is similar to phenomena of the  BL passing through the glass window when the internal $AQN_s$ temperatures could reach enormous values due to much higher density of the soil in comparison with air, see our estimates  in Sect. \ref{sect:glass}.  It explains the hot glassy   rocks found in the area.   It also explains why    no particles of a
green fireball had ever been found, similar to many other UAP reports. It also explains the excess of radiation which often reported as a result of the  UAP impact, similar to BL recorded events discussed in \cite{Zhitnitsky:2025bvy}.  This is because the energy powering the fireballs is the result of annihilation processes of surrounding material with antimatter substance  the fireball made of. 

To conclude this section: the MT events as recorded  by POSS-1 represent the initial stage of the $AQN_s$ evolution when spallation occurs, after which $AQN_s$ starts to propagate in atmosphere. When the velocity of the $AQN_s$ is very high it is likely to emit in UV which is hard to observe by naked  eye. When velocity decreases it starts to emit the visible light, and eventually it hits the ground by heating the surrounding area.    Excess of radiation in broad frequency bands will also remain in the area for sometime. It explicitly shows how MT events and specific type of UAP events as described in Sect \ref{observations} are linked: they simply represent different stages in evolution of the same object.   All estimated temporal and spatial scales are consistent with this proposal (\ref{eq:proposal-MT}).

\section{MT event rate}\label{sec:MT_event_rate} 
The main goal of this section is to argue that our proposal (\ref{eq:proposal-MT}) on the nature of MT events is consistent with statistical analysis carried out in \cite{Villarroel:2021ddh,Villarroel_2025,Villarroel:2025xym}. The basic normalization  for event rate in our proposal   is determined by DM density and galactic velocity along with single   parameter of the model,  the  typical baryon charge $\langle B\rangle \approx 10^{25}$  of the AQN according to (\ref{Phi1}). We shall argue that MT even rate estimated from  (\ref{Phi1}) is consistent with results of refs. \cite{Villarroel:2021ddh,Villarroel_2025,Villarroel:2025xym}. Unfortunately, the direct comparison for fluxes between
theoretical prediction (\ref{Phi1}) and observations  is far from being  straightforward. This is because in papers \cite{Villarroel:2021ddh,Villarroel_2025,Villarroel:2025xym} each MT event counts as independent event, while in our interpretation 
within  AQN framework the MT multiple  events should be treated as the cluster of a single AQN event as explained at the end of Sect. \ref{sect:3}.

We start by estimating   the total number of AQN events for the period 2718 days (7.4 years) which hit the area surrounding  Palomar Observatory from (\ref{Phi1}). The estimation of the effective  area  is very ambiguous problem because spallation may occur at any distance $r$ and altitude $h$ (we estimate the altitude should be   $h \lesssim30 \rm  km$ for air density to be sufficiently high). Therefore, we compute the area for the north hemisphere, which is $2\pi r^2$ and multiply factor 1/2 to account for the night observations only. As a result we arrive from (\ref{Phi1}) to the order of magnitude estimate
\be
\label{Phi2}
N^{\rm AQN}\sim 
  \rm  840~ events \left(\frac{10^{25}}{\langle B\rangle}\right) \left(\frac{r}{30 km}\right)^2 ~~~ ~~~~for ~~~ ~7.4~ years. 
 \ee
This estimate (with few modifications, see below) should not be literally compared   with 83 multiple aligned MT events \cite{Villarroel_2025}. 
This is because MT events are identified with spallation events (not total number of AQN events hitting the area). This introduces 
  the factor analogous  to $\cal{F}$ in estimate for BL events  (\ref{BL-flux1}). We introduce similar suppression factor  $\cal{F}^{\rm MT}$  for MT events which can be estimated from  the argument that MT events strongly correlate with nuclear testing as discussed in Sect. \ref{sect:4}.  In the AQN framework the MT events are always
  almost simultaneous (spatially and temporary) events. 
  Therefore we estimate the number of cluster events when   multiple MT events on the same plate should be counted as a single cluster event as explained in Sect. \ref{sect:3}, i.e.
\be
\label{Phi3}
N^{\rm MT}\sim (N^{\rm AQN}\cdot {\cal{F}}^{\rm MT})\sim
  \rm  80~ events \left(\frac{10^{25}}{\langle B\rangle}\right) \left(\frac{r}{30 km}\right)^2 ~~~~for ~~~{\cal{F}}^{\rm MT}\sim \frac{2\cdot 124}{2718}\approx 0.092,
     \ee
 where factor 124 is number of nuclear testing  days and factor 2 is due to counting two days: day of the test and day after. It is obviously an order of magnitude estimate  at the very   best, as numerical value for ${\cal{F}}^{\rm MT}$ and $r$ are not well known. 
 We still cannot directly compare 83 multiple aligned MT events  from \cite{Villarroel_2025}  with (\ref{Phi3}).   
 This is because   single and double events are not present in  analysis   \cite{Villarroel_2025} by obvious reasons. However, it is expected that the most MT events within AQN framework should be multiple cluster events, such that  83 events used in our comparison as a benchmark value is not very strong   underestimation of the total number of clustering MT events. 
 
 Therefore, we conclude that our estimate (\ref{Phi3}),  as an order of magnitude estimate,  is consistent with results  \cite{Villarroel_2025}. We consider this consistency as a highly nontrivial  result as the basic normalization formula (\ref{Phi1}) depends on SHM   parameters  (DM density and galactic velocity) as well as parameter of the AQN model $\langle B\rangle$ which was extracted by  matching  the  computations in AQN model  with  observed   mysterious phenomena,   such  as  the puzzling  UV excess in our galaxy \cite{Sekatchev:2025ixu}.   It is astonishing that  the normalization factors from dramatically different physics nevertheless produce  the estimate (\ref{Phi3}) which assumes a very reasonable value, being consistent with   the observed results  \cite{Villarroel_2025}. 
 
 One should also add that the same basic normalization formula (\ref{Phi1})  had been previously used to estimate the  BL flux  (\ref{BL-flux1}). Furthermore, the same  
normalization formula (\ref{Phi1}) had been used in estimation of the rate of so-called Telescope Array  (TA) mysterious   bursts\footnote{Telescope Array  (TA) collaboration reported the observations of   mysterious   bursts    when 
   at least three air showers  were recorded within 1 ms which cannot occur with conventional high energy CR. Similar ``Exotic Events"  were observed by AUGER   collaboration. We briefly overview these puzzling events in Sect. \ref{conclusion}. }.  The TA mysterious bursts,  according to the proposal  \cite{Zhitnitsky:2020shd,Liang:2021wjx} is also  the manifestation of the AQN-induced events.   The event rate for similar ``Exotic Events" recorded by  the  AUGER   collaboration 
is also  based on the same normalization formula (\ref{Phi1}) as estimated in \cite{Zhitnitsky:2022swb}. Finally  the frequency of appearance of  two anomalous events    with non-inverted polarity as recorded by 
\textsc{ANITA} collaboration may be also   the   AQN-induced events as argued in \cite{Liang:2021rnv} where the same  basic normalization formula (\ref{Phi1}) had been used. We overview all these puzzling observations  with references  and details in concluding Sect. \ref{conclusion}.  In all our previous applications mentioned above  the event rate was consistent with observations, which further supports our estimate (\ref{Phi3}) in application to the MT clusters.

 %   The basic claim of  \cite{Budker:2020mqk}   supported by the  corresponding estimates  is that the skyquakes are  the direct manifestations of the AQN propagating in atmosphere. 

\section {Concluding comments and Future Developments} \label{conclusion}

     The present work is devoted to   Mysterious Transient events discovered in \cite{2022MNRAS.515.1380S,Villarroel:2019bky,Villarroel:2020knw,Villarroel:2021ddh,2024MNRAS.527.6312S,Villarroel_2025}. Our suggestion on physical nature of these MT puzzling events is formulated as proposal  (\ref{eq:proposal-MT}) and we have presented a number of arguments supporting our proposal in Sect.\ref{sect:proposal} and Sect. \ref{sec:MT_event_rate}, where we argued that all  mysterious features 1.-5. as listed in Sect. \ref{items} find very natural explanation within our proposal treating MT events as AQN-induced events. 
             
             The key element of the proposal is   a  new  DM paradigm when DM is not a weakly interacting massive particle (WIMP) which has been the conventional  paradigm for 40+ years. Instead,  the DM in this new framework, coined as AQN, is very strongly interacting object made of quarks, antiquarks\footnote{We remind the readers that the antimatter in this framework was suggested long ago  \cite{Zhitnitsky:2002qa}, as natural resolution of two fundamental  cosmological puzzles: 1. similarity  between visible and DM components, $\Omega_{\rm DM}\sim \Omega_{\rm visible}$; 2. observed baryon asymmetry of our Universe. These puzzles are automatically resolved in the AQN framework  irrespective to the parameters of the model, see \cite{Zhitnitsky:2021iwg,VanWaerbeke:2026sxs} for review.}    and gluons of SM particle  physics as reviewed in Sect. \ref{AQN}. 
               The AQN behaves as DM object in empty space, but it may produce a number of profound effects when it enters the Earth or other planets or stars where environment is dense. The MT events discovered in \cite{2022MNRAS.515.1380S,Villarroel:2019bky,Villarroel:2020knw,Villarroel:2021ddh,2024MNRAS.527.6312S,Villarroel_2025} is one of such profound manifestations, among many, many others to be briefly reviewed below. It is important to emphasize that all parameters which have been used in our estimates in Sect.\ref{sect:proposal} and Sect. \ref{sec:MT_event_rate} are not ad hoc parameters to fit the observations. Instead, the same parameters have been used previously to study  
                other puzzling observations and which could be naturally explained within   the same AQN framework. 
           Below we briefly overview these mysterious observations to demonstrate the links between MT events  and other   puzzling observations which are close cousins of the MT events.  
             
             \subsection{MT events and their  cousins: from BL events to sky-quakes to ``pseudo-meteorites"}\label{sect:cousins}
             We start with the BL events which was the topic of  \cite{Zhitnitsky:2025bvy} briefly reviewed in Sect. \ref{BL-features}. We have nothing new to add to the description of BL events per se. The only comment we would like to make here is to link the two: MT with BL events.    The spallation event is  identified with MT event according to (\ref{eq:proposal-MT}). After spallation the AQN$_s$ propagates in atmosphere, emits mostly in x rays and UV frequency bands before loosing its enormous initial velocity. After that when AQN$_s$  slows down and propagates at low altitudes it effectively emits light in visible (as well as UV and x rays) frequency bands. At this point 
 it becomes the BL according to the proposal (\ref{eq:proposal}).
                          
             The same AQN$_s$  if they are sufficiently large in size will cross the atmosphere from the formation point, continue to propagate at low altitudes with sufficiently high velocity and eventually hit the ground.  These events are classified as ``pseudo-meteorites", see Sect. \ref{observations} and footnote \ref{footnote:meteorites} for references and details.  This makes a precise link between MT events and ``pseudo-meteorites" events within AQN framework.

    We now describe another mysterious events: the phenomenon known as skyquakes, which have been reported for centuries (similar to BL events) without any viable explanations from the standpoint of conventional physics \cite{skyquakes}.    Skyquakes are extremely rare acoustic events that sound like a cannon shot or a sonic boom coming from the sky
    without any seismic activities, weather related or meteor-related events, which are routinely recorded around the world. It has been proposed  \cite{Budker:2020mqk} that the sky-quakes
    are the AQN  induced events.  There are no much  papers in scientific journals devoted to sky-quakes. However, the are many TV interviews by  meteorologists such as \cite{skyquakes-meteorologist} and by NASA personal  such as \cite{NASA} on the internet. Fortunately, we are aware of a proper recording by a scientific instrument of a single very powerful   event which has been interpreted as a sky-quake event  \cite{Budker:2020mqk}.    This rare event   took place on July 31, 2008, and it  was properly   recorded by the Elginfield Infrasound Array (ELFO) near London, Ontario, Canada \cite{ELFO}. The detection of infrasound was accompanied by nonobservation of any meteors by an all-sky camera network, effectively ruling out a conventional meteor source. Additionally, no meteorites were discovered in the active area. Besides infrasound, seismic impulses were also detected shortly thereafter as ground-coupled acoustic waves around Southwestern Ontario and Northern Michigan. This event was treated as an AQN-induced event by Ref. \cite{Budker:2020mqk}, and it was argued that the energetic properties, infrasound frequency characteristics, and other attributes of the event are consistent with the observed ELFO phenomenon. 
    
     Our previous work  \cite{Budker:2020mqk} had proposed  that the skyquakes are the AQN-induced events, while   in the present work  we  argued that the  MT events are  also AQN-induced events   according to (\ref{eq:proposal-MT}). Therefore, these two naively (observationally) very different phenomena are in fact closely related as they represent different stages in evolution of the same object, the dark matter AQN.      

  Another puzzling observation we want to mention below   could be also the  AQN-induced event  is the observation by the Telescope Array  (TA) collaboration   of  the so-called  mysterious   bursts \cite{Abbasi:2017rvx,Okuda_2019}. The mystery here is that at least three air showers  were recorded within 1 ms (multiple simultaneous events representing the cluster) which cannot occur with conventional high energy Cosmic Rays (CR).   
                We suggested   in   \cite{Zhitnitsky:2020shd} that these puzzling  events  could be related to the AQN-induced events.
   In fact,  in our estimates for the event rate  we used the same formula (\ref{Phi1}) which was used in   our estimates for frequency of the MT events in Sect. \ref{sec:MT_event_rate}. The estimation for the event rate for   TA mysterious   bursts (10 events recorded during 5 years)
   is consistent with our estimation   based on  our normalization for the flux   (\ref{Phi1}).

   Similar, but independent observations had been recorded by another CR laboratory,  the  AUGER   collaboration  \cite{PierreAuger:2021int,2019EPJWC.19703003C,Colalillo:2017uC}.  
    These events (which were coined as  ``Exotic Events") also cannot be explained by canonical CR modelling. At the same time, 
   these ``Exotic Events" recorded by  the  AUGER   collaboration   can be   explained within AQN framework as argued in \cite{Zhitnitsky:2022swb}. Furthermore, the rareness of these ``Exotic Events" (23 events recorded during 13 years) is also explained in the AQN framework by using the same 
    formula (\ref{Phi1}) and it is consistent with counting of the ``Exotic Events" by the  AUGER   collaboration.

    There are many other terrestrial unusual events   when conventional picture cannot explain the observed phenomena. In particular,    the Antarctic Impulsive Transient Antenna  (\textsc{ANITA})  observed      two anomalous events    with non-inverted polarity   \cite{Gorham:2016zah,Gorham:2018ydl}.  Such events correspond to very large inclination angle  when ``something" crossing the earth before exiting from opposite side of earth  at the moment of  recording by \textsc{ANITA}. Such events  are very hard to explain within conventional physics, but could be  explained within AQN framework \cite{Liang:2021rnv}, including the energetic and spectral properties of the radiation.  The estimation for the event rate for   anomalous events   with non-inverted polarity 
   is also consistent with   our formula for the flux   (\ref{Phi1}).

Finally, it has been recently argued in \cite{Zioutas_2020,Argiriou:2025xsq,Zioutas:2026xqb} that numerous enigmatic observations remain challenging to explain within the framework of conventional physics.  In particular, these anomalies include unexpected correlations between temperature variations in the stratosphere and  the total electron content of the Earth's atmosphere
(along with many other mysterious correlations).    It    has been argued in \cite{Zhitnitsky:2024jnk}  that the recorded correlations can be 
generated by the AQN-induced processes. 

The list of mysterious (not yet explained) phenomena in the Earth's atmosphere as listed above is obviously far from being complete.  Nevertheless we opted to collect  
some of them here presenting the  arguments that they might be different faces of the same DM object, the AQN. It gives us a better idea how to  study these (and related)  phenomena in future by designing specific instruments which would account for  physical   manifestations described above.  In fact, the study of historical UAP data    could provide a path for systematic studies of the UAP events and required instruments as recently argued  in \cite{doi:10.1142/S2251171723400068} in context of the Galileo Project or \cite{Szydagis:2024eea} in context of UAPx project. 
    
    \subsection{MT events and future tests}\label{sect:tests}
     Therefore, based on the discussions from previous Sect. \ref{sect:cousins}  we suggest to test our proposal (\ref{eq:proposal-MT})
   relating MT, BL, Pseudo-meteorites, skyquakes    phenomena with AQN-induced effects (representing the DM physics in this framework) as follows. 
   
   First of all, one can search for  infrasound acoustic signals, similar to our suggestion formulated in    \cite{Budker:2020mqk} where we proposed to use Distributed Acoustic Sensing (DAS), which is becoming a conventional tool for seismic and other applications.  The main advantage to use DAS in comparison with other seismic tools is that the DAS is capable of measuring strain changes at all points along the optical fiber at  acoustic, including infrasound, frequencies. This feature  is crucial for our studies of  relatively weak  acoustic waves emitted by  the propagating AQN. If one can install the DAS in the same location where future MT cluster events are planning to be studied one can analyze the correlations between MT cluster  events and acoustic events recorded by DAS which also provides  the directionality of the propagating AQN. The observations of such correlations would definitely discriminate MT cluster  events from any other possible sources such as transients from reflective artificial objects in Earth orbit or solar reflections from artificial surfaces. 
   
  Another suggestion to test our proposal  (\ref{eq:proposal-MT}) is to  install all sky cameras, similar to the ones used in  analysis of the BL spectrum \cite{BL-spectrum-1}, to monitor entire  sky (including very low altitudes)  in the same area where future MT cluster events are planning to be studied.   As we already mentioned previously the propagating  AQN$_s$
  with high velocity at relatively high altitudes are mostly emitting in UV and x ray frequency bands, and therefore cannot be easily observed. This emission pattern changes when  AQN$_s$ slows down at very lower altitudes and starts to emit in visible frequency bands. This is the environment where BL and pseudo-meteorites  had been observed as described in Sect. \ref{sect:cousins}. The analysis of the time correlations between all sky camera and MT events would unambiguously identify the source and nature of these events.

         Last but not least. The LHC machine has been originally designed to discover BSM physics in form of WIMPs, SUSY, and many other elements of 40+years old canonical paradigm. Nothing from this BSM list has been discovered so far. In recent work  \cite{Liang:2026tjs} we suggested to use the   LHC machine  in a dramatically different way which would make the  direct connection  to the topic of the present work, the MT events discovered in \cite{2022MNRAS.515.1380S,Villarroel:2019bky,Villarroel:2020knw,Villarroel:2021ddh,2024MNRAS.527.6312S,Villarroel_2025}.  To be more precise, the source of the MT events are   the AQNs which are always accompanied by the acoustic signal, known as the sky quakes as discussed in Sect.  \ref{sect:cousins}. 
         A similar  acoustic signal     can be recorded at the LHC as suggested in  \cite{Liang:2026tjs} which makes direct connection to MT events because both signals are generated by the {\it same objects}. A brief introduction to this idea is as follows. Since 2010, it has been recognized that the so-called Unidentified Falling Objects (UFOs) may pose a significant limitation to LHC performance \cite{Baer:1379150,Baer:1493018}.   UFOs are commonly attributed to micrometer-sized dust particles released from the beam screen that become attracted to the proton beams and produce beam losses via inelastic proton nucleus collisions \cite{baer_2013_w6p20-zcn10}. However, the mechanism that releases these dust particles remains an open question \cite{Belanger:2020ufo}. 
   %UFO-LHC  events are recorded by beam loss monitors (BLMs) and have been studied in detail, including analyses of multiplicity, time scales, and spatial distribution around the ring. The principal motivation for these studies has been to identify correlations between UFOs and beam parameters in order to improve LHC performance. 
   Despite substantial mitigation efforts in recent years, dust-induced beam losses continue to affect LHC operation \cite{Lechner:2024olj}.
   
  It has been suggested in \cite{Liang:2026tjs} that the  AQN crossing the atmosphere  100 km away from the LHC area generates acoustic waves strong enough to trigger multiple UFO-LHC events. It was suggested to search for correlated multiple UFO-LHC events, which can be easily discriminated from any other sources.  Practically, the LHC can serve as a large broadband acoustic detector for DM in form of the AQNs due to almost perfect vacuum with pressure $\sim 10^{-7}$ Pa inside the beam's pipe. Therefore, the observation  of the acoustic correlated signal   within the LHC ring as described  in \cite{Liang:2026tjs} would unambiguously identify the nature of the DM. We emphasize again that   the source of the UFO-LHC events is originated from the same AQN propagating in atmosphere. In other words, MT cluster events is a different manifestation of the same AQN-induced  events.

         We conclude this work with the following final comment. 
   We advocate an idea that   the MT cluster events discovered in  \cite{2022MNRAS.515.1380S,Villarroel:2019bky,Villarroel:2020knw,Villarroel:2021ddh,2024MNRAS.527.6312S,Villarroel_2025} might be a profound manifestation of the DM physics  when DM  is  made  of (anti)quarks and gluons 
   of the Standard Model as reviewed in Sect.\ref{AQN}.  
  The AQN dark matter model   is  consistent with all presently available cosmological, astrophysical, satellite and ground-based observations.    In fact, it may even shed some light  on the  long standing puzzles and mysteries on Earth as briefly reviewed above in Sect. \ref{sect:cousins} and in cosmos as reviewed in   \cite{Zhitnitsky:2021iwg,VanWaerbeke:2026sxs}.   If validated through the proposed tests and experiments as suggested above it would unambiguously identify the nature of the DM and its role in cosmos and Earth.

               \vspace{0.5cm}

     \section*{Acknowledgements}
      This research was supported in part by the Natural Sciences and Engineering
Research Council of Canada.

  \bibliography{Transients.bib}

\end{document}